\documentclass[12pt]{article}

\usepackage{amssymb}
\usepackage{amsbsy}
\usepackage{amsmath}

\newcommand{\bfm}{\mathbf}
\newcommand{\be}{\begin{equation}} \newcommand{\ee}{\end{equation}}
\newcommand{\bea}{\begin{eqnarray}} \newcommand{\eea}{\end{eqnarray}}
\newcommand{\el}{\nonumber \\}
\newcommand{\re}[1]{(\ref{#1})}
\newcommand{\pat}{\partial}
\newcommand{\adot}{\dot{a}} \newcommand{\addot}{\ddot{a}}
\newcommand{\bdot}{\dot{b}} \newcommand{\bddot}{\ddot{b}}
\newcommand{\ndot}{\dot{n}} 
\newcommand{\Adot}{\dot{A}} \newcommand{\Addot}{\ddot{A}}
\newcommand{\Bdot}{\dot{B}} \newcommand{\Bddot}{\ddot{B}}
\newcommand{\Ndot}{\dot{N}} 
\renewcommand{\Ddot}{\dot{D}} \newcommand{\Dddot}{\ddot{D}}
\newcommand{\Cdot}{\dot{C}} \newcommand{\Cddot}{\ddot{C}}
\newcommand{\Ldot}{\dot{L}} 
\newcommand{\phidot}{\dot{\phi}}
\newcommand{\phiddot}{\ddot{\phi}}

\newcommand{\Ydot}{\dot{Y}} \newcommand{\Yddot}{\ddot{Y}}

\begin{document}

\begin{titlepage}
\begin{flushleft}
       \hfill                      {\tt hep-th/0111279}\\ \hfill
       HIP-2001-64/TH \\ \hfill            December 7, 2001\\
\end{flushleft}
\vspace*{3mm}
\begin{center}
{\Large {\bf On ekpyrotic brane collisions}\\}
\vspace*{12mm} {\large Syksy R\"{a}s\"{a}nen\footnote{Email: syksy.rasanen@helsinki.fi}\\}

\vspace{5mm}

{\em {}Helsinki Institute of Physics \\ P.O. Box 64,
FIN-00014 University of Helsinki, Finland }

\vspace{3mm}

\vspace*{10mm}
\end{center}

\begin{abstract}

\noindent We derive the five-dimensional metrics which describe a non-singular 
boundary brane collision in the ekpyrotic scenario in the context of general
relativity, taking into account brane tension. We show that the metrics
constrain matter created in the collision to have negative energy density or
pressure. In particular, the minimal field content of heterotic M-theory leads
to negative energy density. We also consider bulk brane-boundary brane
collisions and show that the collapse of the fifth dimension is an artifact
of the four-dimensional effective theory.

\end{abstract}

\end{titlepage}

\baselineskip16pt
\setcounter{footnote}{0}

\section{Introduction}

A new cosmological framework called the ekpyrotic scenario has recently
been under intense discussion [1-13].
The scenario has heterotic M-theory \cite{Horava, Lukas} as its origin and
brane cosmology [16-32]
as its context. The setting for the ekpyrotic scenario is an 11-dimensional
spacetime with the topology ${\cal M}_{10}\times S_1/\mathbb{Z}_2$, with
boundary branes at the orbifold fixed points where spacetime terminates. The
boundary branes are called the visible and the hidden brane, with the visible
brane identified with our universe. Six dimensions are compactified on a
Calabi-Yau threefold, leaving the theory effectively five-dimensional.

In the original proposal for a realisation of the ekpyrotic scenario
\cite{Khoury:2001a, Khoury:2001b} a third brane travels from the hidden
brane to collide with the visible brane in an event called ekpyrosis.
Ekpyrosis is posited to transfer some of the energy of this bulk brane onto
the visible brane and thus ignite the big bang at some finite temperature.
A major problem was that during the journey of the bulk brane, the
direction transverse to the branes was contracting, whereas stabilisation
was considered necessary for the post-ekpyrosis era. According to
\cite{Khoury:2001c}, in order to reverse the contraction either the null
energy condition has to be violated or the scale factor of the transverse
direction has to pass through zero. The authors chose not to violate the null
energy condition, and in the second proposal \cite{Khoury:2001c, Khoury:2001d}
there is no bulk brane but the boundary branes themselves collide and then
bounce apart, so that the scale factor passes through zero in what is hoped to
be a non-singular process. This approach has also served as a vital ingredient
in the so-called ``cyclic model of the universe'' \cite{Steinhardt:2001a, Steinhardt:2001b}.

The analysis has been done in the context of a  four-dimensional effective
theory. The evolution of cosmological perturbations
within this framework has been debated, with particular concern about
the matching conditions across the bounce and the validity of the analysis
near the singular point where the scale factor vanishes \cite{Lyth:2001a, Brandenberger:2001, Hwang:2001, Khoury:2001d, Lyth:2001b}.
However, it is not obvious that even the homogeneous and isotropic
background is correctly treated by the four-dimensional effective theory.
It has been observed that the ansatz on which the four-dimensional effective
theory is based cannot support brane matter created by ekpyrosis
\cite{Kallosh:2001b, Enqvist:2001} and does not satisfy the five-dimensional
equations of motion \cite{Kallosh:2001b}, at least with the approximations
made in \cite{Khoury:2001a}. There are also quite general concerns about the
validity of four-dimensional effective theories involving integration over the
transverse direction in brane cosmologies \cite{Mennim:2000, Kallosh:2001b}.

The present paper consists of two main parts. After collecting some necessary
\mbox{equations} in section 2, we study boundary brane collisions with the
full five-dimensional equations in section 3. We derive the
metrics possible under the assumption that the collision is non-singular,
study which of these are ruled out by the field equations and see what are the
constraints on brane matter created by boundary brane ekpyrosis. We compare
with the approach of \cite{Khoury:2001c} and discuss ways to avoid the
constraints on brane matter. In section 4 we study bulk brane-boundary brane
collisions in the moduli space approximation using the five-dimensional
equations. We reassess the collapse of the transverse direction and consider
the validity of the moduli space approximation. In section 5 we summarise our
results and comment on the implications for the ``cyclic model of the
universe''.

\section{The set-up}

\paragraph{The action and the metric.}

The action for both the old and the new ekpyrotic scenario consists of three
parts:
\bea \label{action}
  S = S_{het} + S_{BI} + S_{matter} \ ,
\eea

\noindent where $S_{het}$ is the action of five-dimensional heterotic
M-theory with minimal field content, $S_{BI}$ describes the brane interaction
responsible for brane movement and $S_{matter}$ describes brane matter
created by ekpyrosis.

The simplified action of five-dimensional heterotic M-theory is \cite{Lukas, Khoury:2001a, Kallosh:2001b}
\bea \label{hetaction}
  & & \!\!\!\!\!\! \!\!\!\!\!\! S_{\textrm{het}} \,=\, \frac{M_5^3}{2}\int_{{\cal M}_5} d^5 x \sqrt{-g}\left(R-\frac{1}{2}\pat_A\phi\,\pat^A\phi-\frac{3}{2}\frac{1}{5!}e^{2\phi}{\cal F}_{ABCDE}{\cal F}^{ABCDE}\right) \el
  & & \!\!\!\!\!\! \!\!\!\!\!\! \!\!\!\!\!\! - \sum_{i=1}^3 3\alpha_i M_5^3\int_{{\cal M}_4^{(i)}}\!\! d^4\xi_{(i)}\left(\sqrt{-h_{(i)}}e^{-\phi} 
- \frac{1}{4!}\epsilon^{\mu\nu\kappa\lambda}{\cal A}_{ABCD}\pat_{\mu}X^A_{(i)}\pat_{\nu}X^B_{(i)}\pat_{\kappa}X^C_{(i)}\pat_{\lambda}X^D_{(i)}\right) \ ,
\eea

\noindent where $M_5$ is the Planck mass in five dimensions, $R$ is the
scalar curvature in five dimensions, $e^{\phi}$ is essentially the volume
of the Calabi-Yau threefold and ${\cal A}_{ABCD}$ is a four-form gauge field
with field strength ${\cal F}=d{\cal A}$. The Latin indices run from 0 to 4
and the Greek indices run from 0 to 3. The spacetime is a five-dimensional
manifold ${\cal M}_5= {\cal M}_4\times S_1/Z_2$ with coordinates $x^A$. The
four-dimensional manifolds ${\cal M}_4^{(i)}$, $i=1,2,3$, are the orbifold
planes, called the visible, hidden and bulk branes respectively, with internal
coordinates $\xi^{\mu}_{(i)}$ and tensions $\alpha_i M_5^3$. The tensions are
denoted $\alpha_1=-\alpha$, $\alpha_2=\alpha-\beta$ and $\alpha_3=\beta$.
We leave the sign of $\alpha$ undetermined; the tension of the bulk brane is
always positive, $\beta>0$, and we will assume $\beta<\vert\alpha\vert$.
The tensor $g_{AB}$ is the metric on ${\cal M}_5$ and $h^{(i)}_{\mu\nu}$ are
the induced metrics on ${\cal M}_4^{(i)}$. The functions
$X^A_{(i)}(\xi_{(i)}^{\mu})$ are the coordinates in ${\cal M}_5$ of a point on
${\cal M}_4^{(i)}$ with coordinates $\xi_{(i)}^{\mu}$, in other words they
give the embedding of the branes into spacetime.

The brane interaction term is due to non-perturbative M-theory
effects \cite{Khoury:2001a}. In \cite{Khoury:2001a}, the interaction was given
in the context of a four-dimensional effective action, and it is not known
what it looks like in the five-dimensional picture. However, since the
string coupling is posited to vanish at the brane collision, the contribution
of the brane interaction goes asymptotically to zero before the collision
and rises from zero (or stays zero) after the collision. We will only need
this crucial property for our analysis; the detailed form of the brane
interaction will be unimportant.

Brane matter is assumed to be created in the brane collision. In the old
ekpyrotic scenario, the collision took place between the bulk brane and the
visible brane, so that the hidden brane remained empty. In the new scenario,
the collision is between the boundary branes, so we allow for the possibility
of matter creation on the hidden brane as well. The brane matter action is
\bea \label{matteraction}
  S_{\textrm{matter}} = \sum_{i=1}^2 \int_{{\cal M}_4^{(i)}} d^4\xi_{(i)}\sqrt{-h_{(i)}}{\cal L}_{\textrm{matter}(i)} \ .
\eea

We will consider the following metric ansatz ($t\equiv x^0$, $y\equiv x^4$):
\bea \label{metric}
  ds^2 = -n(t,y)^2 dt^2 + a(t,y)^2 \sum^3_{j=1}(dx^j)^2 + b(t,y)^2 dy^2 \ .
\eea

The branes are taken to be flat and parallel, and we will not consider
brane bending, so the embedding is
\bea
  X^A_{(i)}(\xi_{(i)}^{\mu}) = (t, x^1, x^2, x^3, y_i) \ ,
\eea

\noindent with $y_1=0$, $y_2=R$ and $y_3=Y(t)$, where $R$ is a constant.

\paragraph{The field equations.}

From the action \re{hetaction} with the metric \re{metric} we obtain the
following field equations for ${\cal A}_{ABCD}$ and $\phi$:
\bea \label{eom}
  \square\phi-\frac{3}{5!}e^{2\phi}{\cal F}_{ABCDE}{\cal F}^{ABCDE}
+ \sum_{i=1}^2\delta(y-y_i) b^{-1}6\,\alpha_i e^{-\phi} &=& 0 \el
  D_M (e^{2\phi}{\cal F}^{MABCD}) + \delta_0^{[A}\delta_1^{\ B}\delta_2^{\ C}\delta_3^{\ D]} \sum_{i=1}^2\delta(y-y_i) (-g)^{-1/2}2\,\alpha_i &=& 0 \ ,
\eea

\noindent where $D_M$ is the covariant derivative. The contribution of brane
interaction terms which might couple to ${\cal A}_{ABCD}$ or $\phi$ and
thus affect the equations of motion has been omitted.

\paragraph{The Einstein equation.}

The Einstein equation
\bea
  G_{A B} = \frac{1}{M_5^3} T_{A B}
\eea

\noindent for the action \re{action} and the metric \re{metric} reads
in component form
\bea \label{einstein}
  G^t_{\ t} &=& \frac{3}{b^2} \left[\frac{a''}{a}+\frac{a'}{a}\left(\frac{a'}{a}-\frac{b'}{b}\right)\right] -\frac{3}{n^2}\frac{\adot}{a}\left(\frac{\adot}{a}+\frac{\bdot}{b}\right) \el
  &=& -\frac{1}{4} n^{-2}\phidot^2 - \frac{1}{4} b^{-2}\phi'{}^2 + \frac{3}{4}\frac{1}{5!}e^{2\phi}{\cal F}_{ABCDE}{\cal F}^{ABCDE} \el
  && - \frac{1}{M_5^3}\sum_{i=1}^3 \delta(y-y_i) b^{-1}\rho_{b(i)} + \frac{1}{M_5^3} T^t_{\ t}(BI) \el
  G^j_{\ j} &=& \frac{1}{b^2} \left[2\frac{a''}{a}+\frac{n''}{n}+\frac{a'}{a}
\left(\frac{a'}{a}+2\frac{n'}{n}\right)-\frac{b'}{b}\left(\frac{n'}{n}
+2\frac{a'}{a} \right) \right] \el
  &&-\frac{1}{n^2}\left[2\frac{\addot}{a}+\frac{\bddot}{b}+\frac{\adot}{a}
\left(\frac{\adot}{a}-2\frac{\ndot}{n}\right)
+\frac{\bdot}{b}\left(2\frac{\adot}{a}-\frac{\ndot}{n}\right)\right] \el
  &=& \frac{1}{4} n^{-2}\phidot^2-\frac{1}{4} b^{-2}\phi'{}^2 + \frac{3}{4}\frac{1}{5!}e^{2\phi}{\cal F}_{ABCDE}{\cal F}^{ABCDE} \el
  && + \frac{1}{M_5^3}\sum_{i=1}^3\delta(y-y_i) b^{-1} p_{b(i)} + \frac{1}{M_5^3} T^j_{\ j}(BI)  \el
  G^y_{\ y} &=& \frac{3}{b^2} \frac{a'}{a} \left( \frac{a'}{a} +\frac{n'}{n}\right) - \frac{3}{n^2}\left[\frac{\addot}{a} + \frac{\adot}{a} \left(
\frac{\adot}{a} -\frac{\ndot}{n} \right)\right] \el
  &=& \frac{1}{4} n^{-2}\phidot^2+\frac{1}{4} b^{-2}\phi'{}^2 + \frac{3}{4}\frac{1}{5!}e^{2\phi}{\cal F}_{ABCDE}{\cal F}^{ABCDE} + \frac{1}{M_5^3} T^y_{\ y}(BI) \el
  \label{} G_{ty} &=& 3\left( \frac{n'}{n}\frac{\adot}{a}+\frac{a'}{a}\frac{\bdot}{b}
-\frac{\adot'}{a} \right) = \frac{1}{2}\phidot\, \phi' + \frac{1}{M_5^3} T_{ty}(BI) \ ,
\eea

\noindent where dots and primes stand for derivatives with respect to
$t$ and $y$, respectively, $T_{A B}(BI)$ represents the brane interaction and
$\rho_{b(i)}$ and $p_{b(i)}$ are the energy density and pressure of brane $i$:
\bea \label{rhoandp}
  \rho_{b(i)} &=& \rho_{m(i)} + 3 M_5^3 \alpha_i e^{-\phi} \el
  p_{b(i)} &=& p_{m(i)} - 3 M_5^3 \alpha_i e^{-\phi} \ .
\eea

The terms $\rho_{m(i)}$ and $p_{m(i)}$ are the contribution of brane matter,
present only after ekpyrosis. Note that under the assumption of homogeneity
and isotropy, the energy-momentum tensor of brane matter necessarily has the
ideal fluid form. The delta function part of \re{einstein} reads \cite{Binetruy:1999a}
\bea \label{prejunction}
  3\frac{1}{b}\frac{a''}{a}\bigg|_\delta
&=& -\frac{1}{M_5^3} \sum_{i=1}^2 \delta(y-y_i) \rho_{b(i)} + \mathcal{O}(t) \el
  \frac{1}{b} \left(2\frac{a''}{a}+\frac{n''}{n}\right)\bigg|_\delta
&=& \frac{1}{M_5^3} \sum_{i=1}^2 \delta(y-y_i) p_{b(i)} + \mathcal{O}(t) \ ,
\eea

\noindent where $\mathcal{O}(t)$ stands for possible terms due to a delta
function part in the energy-momentum tensor of the brane interaction. We assume
here and in what follows that terms due to the brane interaction vanish at
least as fast as $t$ near the collision; this does not affect our results in
any way. All that is needed is that the interaction goes smoothly to zero
as the collision is approached. In the bulk brane case there are also
contributions coming from second $t$--derivatives (as well as from mixed
$t$-- and $y$--derivatives) of the metric, but we omit them since we will only
need the junction conditions in the boundary brane brane case. The equations
\re{prejunction} can be rewritten as \cite{Binetruy:1999a}
\bea \label{junction}
  (-1)^{i+1} \frac{1}{b}\frac{n'}{n}\bigg|_{y=y_i}
&=& \frac{1}{6 M_5^3} (2\rho_{b(i)} + 3 p_{b(i)}) + \mathcal{O}(t) \el
 (-1)^{i+1} \frac{1}{b}\frac{a'}{a}\bigg|_{y=y_i}
&=& -\frac{1}{6 M_5^3}\rho_{b(i)} + \mathcal{O}(t) \ .
\eea

\section{Boundary brane collision}

\subsection{Spacetime near the collision}

We will first discuss the new ekpyrotic scenario, where there is no bulk brane
and brane matter is produced in a boundary brane collision. We will not
consider the collision itself, but will concentrate on the periods immediately
before and after the collision. We assume that the behaviour of the model near
the collision can be described by general relativity and classical field
theory, with the equations \re{eom}, \re{einstein}, \re{rhoandp} and
\re{junction}. This obviously requires that there is no curvature singularity
at the collision. The collision problem was studied in \cite{Khoury:2001c},
where it was suggested that the five-dimensional spacetime might behave like a
Milne universe near the collision. We will compare the expectations of
\cite{Khoury:2001c} to our results in section 3.5.

Near the collision, we expand the metric \re{metric} and the size of the
Calabi-Yau threefold as follows:
\bea \label{asymptoticmetric}
  b(t,y) &=& b^{(\pm)}_{k_{(\pm)}}(y) t^{k_{(\pm)}} + \sum_{i={k_{(\pm)}}+1}^\infty b^{(\pm)}_i(y) t^i \qquad\quad t\gtrless0 \el
  n(t,y) &=& n^{(\pm)}_{l_{(\pm)}}(y) t^{l_(\pm)} + \sum_{i=l_{(\pm)}+1}^\infty n^{(\pm)}_i(y) t^i \qquad\quad t\gtrless0 \el
  a(t,y) &=& a^{(\pm)}_{m_{(\pm)}}(y) t^{m_{(\pm)}} + \sum_{i=m_{(\pm)}+1}^\infty a^{(\pm)}_i(y) t^i \quad\quad t\gtrless0 \el
  e^{\phi}(t,y) &=& v^{(\pm)}_0(y) + \sum_{i=1}^\infty v^{(\pm)}_i(y) t^i \qquad\qquad\qquad\quad t\gtrless0 \ ,
\eea

\noindent where $k_{(\pm)}$ are positive constants, $l_{(\pm)}$ and $m_{(\pm)}$
are some constants, $b^{(\pm)}_{k_{(\pm)}}$ are positive functions and
$n^{(\pm)}_{l_{(\pm)}}$ and $a^{(\pm)}_{m_{(\pm)}}$ are non-negative functions
which may have zeros but which do not vanish everywhere. The lower and upper
indices correspond to time before and after the collision, respectively. The
coordinate $t$ is the cosmic time measured on the visible brane, so that
$n(t,0)=1$ by choice of coordinates. The branes have been assumed to bounce
apart instantly, but since we allow the geometry to be discontinuous at the
collision, starting the post-ekpyrosis expansion at $t=0$ rather than at some
$t=t_0>0$ involves no loss of generality. It has been assumed that the volume
of the Calabi-Yau threefold does not grow without bound, since then the
five-dimensional description would certainly break down.

Let us for convenience also define the expansion of $\phi$:
\bea \label{asymptoticphi}
  \phi(t,y) &=& \sum_{i=-\infty}^\infty \phi^{(\pm)}_i(y) t^i \quad\quad t\gtrless0 \ , 
\eea

\noindent where the functions $\phi^{(\pm)}_i$ can be expressed in
terms of $v^{(\pm)}_i$.

In order for the collision to be non-singular, the Riemann tensor
in the local orthonormal basis has to remain bounded as one approaches the
collision (from either side). It then follows from the Einstein equation that
the energy-momentum tensor in the local orthonormal basis also has to remain
bounded. Let us consider first the Riemann tensor and then the energy-momentum
tensor.

\subsection{The Riemann tensor}

In the local orthonormal basis, the nonzero components of the Riemann tensor
in the bulk for the metric \re{metric} are
\bea
  \label{riemann1} R_{\hat t\hat j\hat t\hat j} &=& \frac{1}{b^2} \frac{a'}{a}\frac{n'}{n} - \frac{1}{n^2}\left(\frac{\addot}{a}-\frac{\adot}{a}\frac{\ndot}{n}\right) \\
  \label{riemann2} R_{\hat j\hat j'\hat j\hat j'} &=& - \frac{1}{b^2}\frac{a'{}^2}{a^2} + \frac{1}{n^2}\frac{\adot^2}{a^2} \\
  \label{riemann3} R_{\hat t\hat j\hat y\hat j} &=& \frac{1}{n b}\left( \frac{n'}{n}\frac{\adot}{a}+\frac{a'}{a}\frac{\bdot}{b}-\frac{\adot'}{a}\right)  \\
  \label{riemann4} R_{\hat t\hat y\hat t\hat y} &=& \frac{1}{b^2} \left(\frac{n''}{n}-\frac{b'}{b}\frac{n'}{n}\right) - \frac{1}{n^2}\left(\frac{\bddot}{b}-\frac{\bdot}{b}\frac{\ndot}{n}\right) \\
  \label{riemann5} R_{\hat j\hat y\hat j\hat y} &=& - \frac{1}{b^2} \left(\frac{a''}{a} - \frac{b'}{b}\frac{a'}{a}\right) + \frac{1}{n^2}\frac{\adot}{a}\frac{\bdot}{b} \ ,
\eea

\noindent where $j$ and $j'\neq j$ are spatial directions parallel to the
brane.  The Riemann tensor on the brane is\footnote{Note that this is not
the same as the five-dimensional bulk Riemann tensor evaluated at the brane
position.}
\bea
  ^{(i)} R_{\hat t\hat j\hat t\hat j} &=& - \frac{1}{n^2}\left(\frac{\addot}{a}-\frac{\adot}{a}\frac{\ndot}{n}\right)\bigg|_{y=y_i} \el
  ^{(i)} R_{\hat j\hat j'\hat j\hat j'} &=& \frac{1}{n^2}\frac{\adot^2}{a^2}\bigg|_{y=y_i} \ ,
\eea

\noindent where the index $i$ refers to the four-dimensional quantity
measured on brane $i$. Concentrating on the visible brane, we have
\bea \label{riemannbrane}
  ^{(1)} R_{\hat t\hat j\hat t\hat j} &=& -\frac{\addot}{a}\bigg|_{y=0} \el
  ^{(1)} R_{\hat j\hat j'\hat j\hat j'} &=& \frac{\adot^2}{a^2}\bigg|_{y=0} \ .
\eea

\paragraph{The boundedness requirement.}

Let us first consider the boundedness of the Riemann tensor on the brane.
From \re{riemannbrane} we see that if the scale factor on the visible brane
approaches zero or diverges, the Riemann tensor on the brane grows without
bound\footnote{The same conclusion can also be obtained from the Riemann
tensor in the bulk with the help of the junction conditions \re{junction},
assuming that the brane energy density and pressure remain bounded.}.
We conclude that the scale factor on the visible brane approaches a finite
value as the branes approach each other.

Let us now turn to the Riemann tensor in the bulk. Since the calculation is the
same before and after the collision, we temporarily drop the index $(\pm)$.
We will denote the first terms in the series expansion \re{asymptoticmetric}
of $n$ and $a$ whose $y$--derivative does not vanish everywhere by
$n_{\tilde l}$ and $a_{\tilde m}$. (That is, $n'_i(y)=0 \ \forall\ i<\tilde l$,
$a'_i(y)=0 \ \forall\ i<\tilde m$.) With the series expansion
\re{asymptoticmetric}, the leading terms of the Riemann tensor
\re{riemann1}--\re{riemann5} read
\bea
  \label{riemann1as} R_{\hat t\hat j\hat t\hat j} &\simeq& t^{-2 k+\tilde l-l+\tilde m-m}\frac{1}{b_k^2} \frac{a'_{\tilde m}}{a_m}\frac{n'_{\tilde l}}{n_l}
- t^{-2 l-2} \frac{1}{n_l^2} m (m-l-1) \\
  \label{riemann2as} R_{\hat j\hat j'\hat j\hat j'} &\simeq& - t^{-2 k + 2\tilde m -2 m}\frac{1}{b_k^2}\frac{a'{}_{\tilde m}^2}{a_m^2} + t^{-2 l-2} \frac{1}{n_l^2} m^2 \\
  \label{riemann3as} R_{\hat t\hat j\hat y\hat j} &\simeq& t^{-k-l-1}\frac{1}{n_l b_k}\left( t^{\tilde l-l}m\frac{n'_{\tilde l}}{n_l}+t^{\tilde m-m}(k-\tilde m)\frac{a'_{\tilde m}}{a_m} \right)  \\
  \label{riemann4as} R_{\hat t\hat y\hat t\hat y} &\simeq& t^{-2k+\tilde l -l}\frac{1}{b_k^2} \left(\frac{n''_{\tilde l}}{n_l}-\frac{b_k'}{b_k}\frac{n'_{\tilde l}}{n_l}\right) - t^{-2 l-2} \frac{1}{n_l^2} k (k-l-1) \\
  \label{riemann5as} R_{\hat j\hat y\hat j\hat y} &\simeq& - t^{-2k+\tilde m -m}\frac{1}{b_k^2} \left(\frac{a''_{\tilde m}}{a_m} - \frac{b'_k}{b_k}\frac{a'_{\tilde m}}{a_m}\right) + t^{-2 l-2} \frac{1}{n_l^2} m k \ .
\eea

Recall that we have $n(t,0)=1$ by choice of coordinates, and that according to
\re{riemannbrane} $a(t,0)\simeq$ finite constant. This is only possible if
$l,m\le0$ and $n_i(0)=0\ \forall\ i\neq0$, $a_i(0)=0\ \forall\ i<0$. Note
that the coefficients only have to vanish at the visible brane; it is possible
for them to be non-zero elsewhere. In particular, if $l<0$, then $n_l$
must be zero at the visible brane, but not everywhere. This of course
means that $n'_l$ does not vanish identically, so that
$\tilde l=l$. Similarly, $m<0$ implies $\tilde m=m$. Also, finiteness of the
brane energy density and pressure imply, via the junction conditions
\re{junction},
\bea \label{braneconstraint}
  n'_i(0) &=& a'_i(0) = 0 \quad\quad \forall\ i<k \\
  n'_k(0) &\neq& 0, \ a'_k(0) \neq 0 \ ,
\eea

\noindent which obviously means $\tilde l, \tilde m\le k$. Note that since
$l$ and $m$ have turned out to be integers, $k$ must also be an integer.

Let us assume that $l<0$. Then the first term of \re{riemann1as} is
proportional to $\vert t\vert^{-2 k+\tilde m-m}\ge \vert t\vert^{-k}$
with a non-vanishing coefficient, and thus divergent. Since the second term is
proportional to at worst
$\vert t\vert^{-2l-2}=\vert t\vert^{2\vert l\vert -2}\le \vert t\vert^0$,  it
is bounded and cannot cancel the divergence of the first term. So, we must
have $l=0$.

Let us now assume that $m<0$. Then, in order for it to be possible for
the divergent terms in \re{riemann1as} and \re{riemann2as} to cancel, we must
have $k=1$ and $\tilde l=0$. But then the first term in \re{riemann4as}
is more divergent than the second, and thus its coefficient must vanish,
yielding $n'_0/b_k=$ constant. However, according to \re{braneconstraint}
we must have $n'_0(0)=0$, implying that $n'_0$ vanishes everywhere, in
contradiction with the definition of $n_{\tilde l}$. We conclude that $m=0$.

Given $l=m=0$, it follows straightforwardly from \re{riemann1as},
\re{riemann2as} and \re{riemann3as} that $\tilde l=\tilde m=k$. The
first three components of the Riemann tensor provide no further insight.
The results of the remaining two equations depend on the value of $k$,
so let us consider the different possibilities separately.

\paragraph{$\bfm{k=1}$.}

For the simplest possibility, the cancellation of the divergences in
\re{riemann4as} and \re{riemann5as} is equivalent to the following equations
\bea
  \frac{1}{b_k^2} \left(n''_1 - \frac{b_k'}{b_k}n'_1\right) - n_1 &=& 0 \el
  \frac{1}{b_k^2} \left(a''_1 - \frac{b'_k}{b_k}a'_1\right) - a_1 &=& 0 \ .
\eea

With the coordinate choice $b_k(y)=B$, with $B$ a positive constant,
the above equations reduce to
\bea
  n''_1 - B^2 n_1 &=& 0 \el
  a''_1 - B^2 a_1 &=& 0 \ ,
\eea

\noindent with the solutions
\bea
  n_1(y) &=& N_1 \sinh(B y) \el
  a_1(y) &=& A_1 \sinh(B y) + \tilde A_1 \cosh(B y) \ ,
\eea

\noindent where $N_1$ and $A_1$ are non-zero constants, $\tilde A_1$ is a
constant which may be zero and we have taken into account $n_1(0)=0$. The
metric \re{metric} near the collision is
\bea \label{asymptotick=1}
  b(t,y) &=& B t + \mathcal{O}(t^2) \el
  n(t,y) &=& 1 + N_1 \sinh(B y) t + \mathcal{O}(t^2) \el
  a(t,y) &=& 1 + \left( A_1 \sinh(B y) + \tilde A_1 \cosh(B y) \right) t + \mathcal{O}(t^2) \ ,
\eea

\noindent where we have set $A_0=1$.

\paragraph{$\bfm{k\ge2}$.}

Now the cancellation of the leading divergences in \re{riemann4as} and
\re{riemann5as} is equivalent to the equations
\bea
  \frac{1}{b_k^2} \left(n''_k - \frac{b_k'}{b_k}n'_k\right) - 2\delta_{2k} &=& 0 \el
  \frac{1}{b_k^2} \left(a''_k - \frac{b'_k}{b_k}a'_k\right) &=& 0 \ ,
\eea

\noindent where $\delta_{2k}$ is the Kronecker delta. The above equations have
the solution
\bea \label{nandak}
  \frac{1}{b_k} n'_k &=& N_k + 2 \delta_{2k}\int_0^y dz b_k(z) \el
  \frac{1}{b_k} a'_k &=& A_k \ ,
\eea

\noindent where $N_k$ and $A_k$ are non-zero constants. With the metric choice
$b_k(y)=B$ the solutions reduce to
\bea
  n_k(y) &=& N_k B y + \delta_{2k} B^2 y^2 \el
  a_k(y) &=& A_k B y + \tilde A_k \ ,
\eea

\noindent where $\tilde A_k$ is a constant.

The cancellation of subleading divergences in \re{riemann4as} and
\re{riemann5as} imposes $k-1$ relations between the higher order coefficients
$n_{k+i}$ and $a_{k+i}$. To leading order, the metric \re{metric} is
\bea \label{asymptotickge2}
  b(t,y) &=& B t^k + \mathcal{O}(t^{k+1}) \el
  n(t,y) &=& 1 + ( N_k B y + \delta_{2k} B^2 y^2 ) t^k + \mathcal{O}(t^{k+1}) \el
  a(t,y) &=& 1 + \sum_{i=1}^{k-1} A_i t^i + ( A_k B y + \tilde A_k ) t^k + \mathcal{O}(t^{k+1}) \ ,
\eea

\noindent where $A_i$ are constants which may be zero, and we have set $A_0=1$.

\paragraph{The no-flow requirement.}

In addition to the boundedness requirement, there is another constraint the
metric should satisfy: energy should not flow off spacetime at the branes,
\bea \label{noflowmetric}
  G_{\hat t\hat y}\bigg|_{y=y_i} = 0 \ .
\eea

The condition \re{noflowmetric} is satisfied to leading order by virtue
of the component \re{riemann3} of the Riemann tensor being bounded.
Subleading terms of \re{noflowmetric} involve higher order coefficients
$n_{k+i}$ and $a_{k+i}$. As noted, the boundedness of the components
\re{riemann4} and \re{riemann5} of the Riemann tensor imposes $k-1$ conditions
on the same coefficients. However, the boundedness and no-flow conditions are
compatible and can all be simultaneously satisfied.

\subsection{The energy-momentum tensor}

We have derived the metrics allowed by the non-singularity and
non-boundedness conditions of the Riemann tensor. Let us now see which of
these metrics, \re{asymptotick=1} and \re{asymptotickge2}, are allowed by the
same conditions of the energy-momentum tensor.

In the local orthonormal basis, the bulk energy-momentum tensor given in
\re{einstein} is (after eliminating ${\cal A}_{ABCD}$ by using its equation
of motion)
\bea
  \label{emtensortt} \frac{1}{M_5^3} T_{\hat t\hat t} &=& \frac{1}{4} n^{-2}\phidot^2 + \frac{1}{4} b^{-2}\phi'{}^2 + \frac{3}{4}\alpha^2 e^{-2\phi} + \frac{1}{M_5^3} T_{\hat t\hat t}(BI) \\
  \frac{1}{M_5^3} T_{\hat j\hat j} &=& \frac{1}{4} n^{-2}\phidot^2-\frac{1}{4} b^{-2}\phi'{}^2-\frac{3}{4}\alpha^2 e^{-2\phi} + \frac{1}{M_5^3} T_{\hat j\hat j}(BI) \\
  \label{emtensoryy} \frac{1}{M_5^3} T_{\hat y\hat y} &=& \frac{1}{4} n^{-2}\phidot^2+\frac{1}{4} b^{-2}\phi'{}^2-\frac{3}{4}\alpha^2 e^{-2\phi} + \frac{1}{M_5^3} T_{\hat y\hat y}(BI) \\
  \label{emtensorty} \frac{1}{M_5^3} T_{\hat t \hat y} &=& \frac{1}{2} n^{-1} b^{-1} \phidot\, \phi' + \frac{1}{M_5^3} T_{\hat t\hat y}(BI) \ ,
\eea

\noindent and the brane energy-momentum tensor given in \re{einstein} and
\re{rhoandp} is
\bea
  ^{(i)} T_{\hat t\hat t} &=& \rho_{m(i)} + 3 M_5^3 \alpha_i e^{-\phi} + ^{(i)}\! T_{\hat t\hat t}(BI) \el
  ^{(i)} T_{\hat j\hat j} &=& p_{m(i)} - 3 M_5^3 \alpha_i e^{-\phi} + ^{(i)}\! T_{\hat j\hat j}(BI) \ ,
\eea

\noindent where the index $i$ refers to the four-dimensional quantity
measured on brane $i$. Since there is no bulk brane, $\alpha_i=(-1)^i \alpha$.

\paragraph{The boundedness requirement.}

Requiring the bulk energy-momentum tensor to remain bounded gives the
conditions
\bea
  \label{phicond1} n^{-1}\phidot &=& \mathcal{O}(t^0) \\
  \label{phicond2} b^{-1}\phi' &=& \mathcal{O}(t^0) \\
  \label{phicond3} e^{-\phi} &=& \mathcal{O}(t^0) \ .
\eea

The boundedness of the brane energy-momentum tensor does not impose any
additional constraints. In terms of the series expressions
\re{asymptoticmetric} and \re{asymptoticphi}, the conditions
\re{phicond1}--\re{phicond3} read, given $l=0$,
\bea
  \label{phicond1b} \phi_i(y) &=& 0 \quad\quad \forall \ i<0 \\
  \label{phicond2b} \phi_i'(y) &=& 0 \quad\quad \forall \ i<k \\
  \label{phicond3b} \label{volume0} v_0(y) &\neq& 0 \ .
\eea

\paragraph{The no-flow requirement.}

In addition to the boundedness requirement, we should again require that
energy does not flow away from spacetime at the branes,
\bea
  T_{\hat t\hat y}\bigg|_{y=y_i} = 0 \ .
\eea

With \re{emtensorty} this gives, since $b^{-1}\phi'$ is finite at the
branes due to the equation of motion \re{eomdeltab},
\bea \label{prenoflow}
  n^{-1}\phidot\bigg|_{y=y_i} \longrightarrow^{\!\!\!\!\!\!\!\!\!\!\!\!\! t\rightarrow0^{\pm}} 0 \ ,
\eea

\noindent a slightly but crucially stronger condition than \re{phicond1}. In
terms of the series expression \re{asymptoticphi}, the condition
\re{prenoflow} reads
\bea \label{noflow}
  \phi_1(y_i) = 0 \ .
\eea

\paragraph{Newton's constants.}

There has been some concern \cite{Khoury:2001c, Steinhardt:2001b}
that a vanishing transverse direction leads to a divergent Newton's constant
and therefore large quantum fluctuations. However, the calculations
have dealt with a Newton's constant in a four-dimensional effective theory.
The Newton's constant measured in the bulk of the five-dimensional theory is
of course constant, while the Newton's constant measured on a brane is
$\alpha_i e^{-\phi}/(16\pi M_5^3)$, where $\alpha_i$ is the brane tension and
$e^{-\phi}$ is evaluated at the brane position
\cite{Mennim:2000, Enqvist:2001} . As long as the size of the Calabi-Yau
threefold stays finite the Newton's constant(s) in the five-dimensional theory
are completely well-behaved, regardless of the behaviour of the transverse
direction.

\paragraph{The field equation.}

There is one more condition that the energy-momentum tensor should satisfy:
covariant conservation. In the case of the metric, covariant
conservation (of the Einstein tensor) is an identity, but for the
energy-momentum tensor it provides a non-trivial constraint. The covariant
conservation law of the energy-momentum tensor is in this case (after
eliminating ${\cal A}_{ABCD}$ by using its equation of motion) equivalent to
the equation of motion of $\phi$,
\bea
  \label{eomb} - n^{-2} \left[\phiddot + \left(-\frac{\ndot}{n}+3\frac{\adot}{a}+\frac{\bdot}{b}\right)\phidot\right] \el
  + b^{-2} \left[\phi'' + \left(\frac{n'}{n}+3\frac{a'}{a}-\frac{b'}{b}\right)\phi'\right] + 3\,\alpha^2 e^{-2\phi} &=& \mathcal{O}(t^{-k+1}) \\
  \label{eomdeltab} \delta(y-y_i) (b^{-1}\phi'-3\alpha e^{-\phi}) &=& \mathcal{O}(t) \ ,
\eea

\noindent where the right-hand sides are the possible contribution of the brane
interaction. The brane part \re{eomdeltab} of the field equation is satisfied
provided that
\bea \label{asymptoticeomdelta}
  \frac{1}{b_k(y_i)}\phi_k'(y_i) &=& 3\alpha e^{-\phi_0} \ .
\eea

Inserting the expansions \re{asymptoticmetric} and \re{asymptoticphi} into the
bulk part \re{eomb} of the field equation and taking into account
\re{phicond1}--\re{phicond3} and the previous section's results
$l=m=0$, $\tilde l=\tilde m=k$,
the leading terms are
\bea \label{asymptoticeom}
  -k t^{-1} \phi_1 + b_k^{-2} t^{-k} \left(\phi''_k -\frac{b'_k}{b_k}\phi'_k\right) &=& \mathcal{O}(t^{-k+1}) \ .
\eea

Let us consider different values of $k$ separately.

\paragraph{$\bfm{k=1}$.}

With the coordinate choice $b_k(y)=B$, \re{asymptoticeom} simplifies to
\bea
  \phi''_1 - B^2 \phi_1 &=& 0 \ ,
\eea

\noindent with the familiar solution 
$\phi_1(y)=\lambda \cosh(B y) + \tilde\lambda \sinh (B y)$, where
$\lambda, \tilde\lambda$ are constants. The requirement that no energy flows
away from spacetime, \re{noflow} leads to $\phi_1(y)=0$. But according to 
\re{asymptoticeomdelta}, we should have $\phi'_k(y_i)\neq0$. We conclude that
$k=1$ is ruled out by the no-flow condition of $\phi$. 

\paragraph{$\bfm{k\ge2}$.}

In this case the leading term of \re{asymptoticeom} gives
\bea
  \phi''_k -\frac{b'_k}{b_k}\phi'_k &=& 0 \ ,
\eea

\noindent the solution of which is also familiar,
\bea \label{phik}
  \frac{1}{b_k}\phi'_k &=& \textrm{constant} \el
  &=& 3 \alpha e^{-\phi_0} \ ,
\eea

\noindent where we have on the second line used \re{asymptoticeomdelta}. With
the coordinate choice $b_k(y)=B$ we have
\bea
  \phi_k(y) &=& 3 \alpha e^{-\phi_0} B y + \varphi_k \ ,
\eea

\noindent where $\varphi_k$ is a constant. The subleading terms may involve
the brane interaction and thus cannot provide any information.

\subsection{Constraints on brane matter}

We have derived the metrics allowed by the boundedness and no-flow conditions
of the Riemann tensor, \re{asymptotick=1} and \re{asymptotickge2}. We have
then seen that the same conditions of the minimal energy-momentum tensor
of heterotic M-theory metric allow only the metric \re{asymptotickge2}.
Let us now consider what the metrics \re{asymptotick=1} and \re{asymptotickge2}
have to say on the issue of brane matter, setting for a moment aside the
constraint due to the energy-momentum tensor.

It is a known feature of brane cosmologies that limitations on brane matter
may arise in constrained metric configurations, most notably those with
a factorisable metric \cite{Mohapatra:2000, Lesgourges:2000} or $\bdot=0$
\cite{Grinstein:2000, Enqvist:2000, Binetruy:2001}. The near-collision
metrics \re{asymptotick=1} and \re{asymptotickge2} do not fall into either
class, but they do have quite a restrictive form. Since we need to discuss the
pre- and post-ekpyrosis eras separately, we return the index $(\pm)$.

\paragraph{$\bfm{k=1}$.}

Putting together the junction conditions \re{junction}, the constraints
\re{phicond2b} and \re{phicond3b} on $\phi$, and the metric
\re{asymptotick=1}, we have
\bea
  \label{constraintk=11} && N_1^{(\pm)} \cosh(B^{(\pm)} y_i) \el
  && = \frac{1}{2} \alpha e^{-\phi_0} - \theta(t)\frac{1}{6 M_5^3}(-1)^i\left(2\rho_{m(i)}(0) + 3 p_{m(i)}(0)\right) \\
  \label{constraintk=12} && A_1^{(\pm)}\cosh(B^{(\pm)} y_i)+\tilde A_1^{(\pm)}\sinh(B^{(\pm)} y_i) \el
  && = \frac{1}{2} \alpha e^{-\phi_0} + \theta(t)\frac{1}{6 M_5^3}(-1)^i\rho_{m(i)}(0) \ ,
\eea

\noindent where $\theta(t)$ is the step function,
$\rho_{m(i)}(0)\equiv\rho_{m(i)}(t=0)$ is the energy density of matter
created on brane $i$ by ekpyrosis, and $p_{m(i)}(0)$ is the corresponding
pressure. In general, we of course cannot set the two functions
$b_{k_{(\pm)}}^{(\pm)}(y)$ to a constant simultaneously both before and after
ekpyrosis, so it should be understood that we are using a different
$y$--coordinate for the pre- and post-ekpyrosis eras.

For the pre-ekpyrosis era, \re{constraintk=11} cannot be satisfied since
the \emph{l.h.s.} is different at different branes whereas  $\phi_0$ is
constant due to the boundedness requirement \re{phicond2b}. We conclude that
$k_{(-)}=1$ is excluded.

For the post-ekpyrosis era, the junction conditions \re{constraintk=11}
and \re{constraintk=12} read
\bea
  6 M_5^3 N_1^{(+)} &=& 3 \alpha M_5^3 e^{-\phi_0} + 2\rho_{m(1)}(0) + 3 p_{m(1)}(0) \el
  6 M_5^3 N_1^{(+)} \cosh(B^{(+)} R) 
  &=& 3 \alpha M_5^3 e^{-\phi_0} - 2\rho_{m(2)}(0) - 3 p_{m(2)}(0) \el
  6 M_5^3 A_1^{(+)} &=& 3 \alpha M_5^3 e^{-\phi_0} - \rho_{m(1)}(0) \el
  6 M_5^3 \left( A_1^{(+)}\cosh(B^{(+)} R)+\tilde A_1^{(+)}\sinh(B^{(+)} R) \right) &=& 3 \alpha M_5^3 e^{-\phi_0} + \rho_{m(2)}(0) \ .
\eea

The last two equations simply give $A_1$ and $\tilde A_1$ in terms of
$\rho_{m(1)}(0)$ and $\rho_{m(2)}(0)$. However, the first two equations
provide the constraint
\bea \label{velocity}
  \frac{3 \alpha M_5^3 e^{-\phi_0} - 2\rho_{m(2)}(0) - 3 p_{m(2)}(0)}{3 \alpha M_5^3 e^{-\phi_0} + 2\rho_{m(1)}(0) + 3 p_{m(1)}(0)} &=& \cosh(B^{(+)} R) > 1 \ ,
\eea

\noindent which implies
\bea
  2\rho_{m(1)}(0) + 3 p_{m(1)}(0) + 2\rho_{m(2)}(0) + 3 p_{m(2)}(0) < 0 \ .
\eea

If we want to avoid negative energy densities we will inevitably have negative
pressures. Having matter with positive energy density on the negative tension
brane may be problematic, since the Newton's constant on a brane is
proportional to the brane tension \cite{Mennim:2000, Enqvist:2001}. For the
standard four-dimensional Hubble law $H^2=8\pi G_N\rho_m/3$, it would
be impossible for $G_N\rho_m$ to be negative,
but in the ekpyrotic scenario it may be possible, depending on the
exact form of the Hubble law on the brane \cite{Enqvist:2001}. In the simple
case that the hidden brane is empty, matter on the visible brane must satisfy
$p_{m(1)}(0)<-2/3\rho_{m(1)}(0)$. In any case, matter on at least one brane
will not be just radiation but something more exotic.

Note that \re{velocity} relates the post-ekpyrosis expansion velocity
$B^{(+)} R$ directly to the brane energy densities and pressures, and thus to
the ekpyrotic temperature. For ekpyrotic temperatures much smaller than the
scale of the brane tension (which is reasonably of the order of the Planck
scale), the velocity $B^{(+)} R$ will be small, of the order of
$T^2/(\vert\alpha\vert M_5^3 e^{-\phi_0})^{1/2}\sim T^2 M_4/M_5^3$, where
$T$ is the ekpyrotic temperature and $M_4$ is the Planck mass on the visible
brane, and we have taken into account the relation
$\vert\alpha\vert e^{-\phi_0}=16\pi M_5^3/M_4^2$
\cite{Mennim:2000, Enqvist:2001}.

\paragraph{$\bfm{k\ge2}$.}

Putting together the junction conditions \re{junction}, the constraints
\re{phicond2b} and \re{phicond3b} on $\phi$ and the metric \re{asymptotickge2},
we have
\bea \label{matterkge2}
  && \!\!\!\!\!\!\!\!\!\! A_{k_{(\pm)}} = \frac{1}{2 M_5^3} \alpha e^{-\phi_0} + \theta(t)\frac{1}{6 M_5^3}(-1)^i\rho_{m(i)}(0) \el
  && \!\!\!\!\!\!\!\!\!\! N_{k_{(\pm)}} + 2 \delta_{2k_{(\pm)}}\int_0^{y_i}\!\!\!\! dz b_{k_{(\pm)}}^{(\pm)}(z) = \frac{1}{2 M_5^3} \alpha e^{-\phi_0} - \theta(t)\frac{1}{6 M_5^3}(-1)^i\left(2\rho_{m(i)}(0) + 3 p_{m(i)}(0)\right) \ .
\eea

For the pre-ekpyrosis era we just obtain the requirement $k_{(-)}\ge3$. For
the post-ekpyrosis era we obtain the following constraints on brane matter
\bea
  \rho_{m(1)}(0) &=& -\rho_{m(2)}(0) \el
  p_{m(1)}(0) &=& -p_{m(2)}(0) - 4 M_5^3\delta_{2k_{(+)}}\int_0^R\!\!\!\! dz b_{k_{(+)}}^{(+)}(z) \ .
\eea

If the energy density and pressure of matter created on the visible brane is
positive, a corresponding negative energy density, along with negative
pressure has to be created on the hidden brane. So, matter on at least one of
the branes violates the null energy condition\footnote{Except in the trivial
case $\rho_{m(i)}(0)+p_{m(i)}(0)=0$, possible for $k_{(+)}\ge3$.}. We conclude
that $k_{(+)}\ge2$ is excluded by the null energy condition.

\subsection{Comparison with the Milne metric}

A preliminary investigation into how a brane collision which looks singular
in a four-dimensional effective theory might be well-behaved in five
dimensions was conducted in \cite{Khoury:2001c}. It was assumed that near the
collision one can neglect the tensions of the branes and approximate the
five-dimensional spacetime with a compactified Minkowski metric. (A part of)
the Minkowski spacetime can be written in Milne coordinates, so that it looks
as follows:
\bea \label{milnemetric}
  ds^2 = -dt^2 + \sum^3_{j=1}(dx^j)^2 + B^2 t^2 dy^2 \ ,
\eea

\noindent where $B$ is a positive constant. This approximation essentially
consists of the following assumptions: near the collision i) the
time-dependence of the metric coefficient $b$ is linear, ii) the
time-dependence of the metric coefficients $n$ and $a$ is of higher order
than that of $b$ and iii) the $y$--dependence in the metric coefficients can
be neglected.

We have now studied the brane collision with the five-dimensional equations,
under the assumption that the five-dimensional theory is non-singular. The
metric near a non-singular collision is, according to \re{asymptotick=1} and
\re{asymptotickge2},
\bea \label{realmetric}
  ds^2 &\simeq& - ( 1+ n_{k_{(\pm)}}(y) t^{k_{\pm}} )^2 dt^2 + ( 1 + \sum_{i=1}^{k_{(\pm)}-1} A_i t^i + a_{k_{(\pm)}}(y) t^{k_{\pm}} )^2 \sum^3_{j=1}(dx^j)^2 \el
  && + B^{(\pm)^2} t^{2 k_{\pm}} dy^2 \ ,
\eea

\noindent where we have set $b_{k_{(\pm)}}^{(\pm)}=B^{(\pm)}$, and
$n_{k_{(\pm)}}$ and $a_{k_{(\pm)}}$ are given in \re{asymptotick=1} and
\re{asymptotickge2}, and $k_{(-)}\ge3$, $k_{(+)}\ge1$. The constraints on
brane matter \re{matterkge2} further say that in order to avoid negative
energy densities we should have $k_{(+)}=1$. The above metric shows
that before the collision $b$ vanishes at least as fast as $t^3$, while after
the  collision it can vanish like $t$ or like some higher power, that the
time-dependence of $n$ and $a$ is at least of the same order as that of $b$,
and that the $y$--dependence of $n$ and $a$ cannot be neglected.  One can
confirm from the Riemann tensor \re{riemann1}-\re{riemann5} that the $t$-- and
$y$--dependence of $n$ and $a$ do make a significant contribution to
physical quantities arbitrarily near the collision.

The physical reason for the failure of the approximation \re{milnemetric}
is the presence of brane tension (and brane matter). The Milne metric
\re{milnemetric} describes a spacetime with no curvature, but
the calculation leading to the true metric \re{realmetric} shows that
energy density on the brane will always curve spacetime in a manner that
cannot be ignored; this is quite transparent in \re{constraintk=11},
\re{constraintk=12} and \re{matterkge2}, which show that $n_{k_{(\pm)}}$ and
(up to an additive \mbox{constant)} $a_{k_{(\pm)}}$ are proportional to brane
energy density and pressure. Were the brane tension turned off, the Milne
metric \re{milnemetric} could be a good approximation.

\subsection{Discussion on boundary brane collisions}

\paragraph{Summary.}

We have derived the metrics which are possible under the assumption
that the five-dimensional description remains valid, there is no curvature
singularity and the brane energy density remains finite. We have then
shown which of these metrics are allowed by the non-singularity and no-flow
conditions of the energy-momentum tensor, and what are the
constraints on brane matter. For the pre-ekpyrosis era everything works out,
provided that the transverse direction vanishes at least as fast as $t^3$.
However, for the post-ekpyrosis era, the single possibility that would avoid
negative energy densities, the transverse direction vanishing like $t$, is
excluded by the no-flow condition of the energy-momentum tensor.

\paragraph{Ways out.}

The rather forbidding conclusions on negative energy densities and/or
pressures have been obtained in the context of five-dimensional heterotic 
M-theory with minimal field content and with dynamics dictated by
general relativity and classical field theory. The relevant question is now
which way the investigation should be generalised in order to avoid the
unwanted results.

A simple remedy might be to turn on a new field in the five-dimensional
action. The problem of negative energy density is solved if the energy flow
associated with this new field compensates for the energy flow of $\phi$ at
the boundary of spacetime, so that the no-flow condition does not imply
$\phidot=0$ at the branes and thus exclude $k_{(+)}=1$. However, the scenario
would still be left with the problem of converting the negative pressure brane
matter into positive pressure radiation and dealing with the inflation
possibly onset by the negative pressure\footnote{When $\phidot\neq0$ at the
visible brane, the contribution of $\phidot$ may decelerate the universe so
much that even a large negative pressure does not lead to inflation. For
details on the effects of $\phidot$ on cosmology on the visible brane, see
\cite{Enqvist:2001}.}. It would be desirable to avoid the exotic
matter altogether.

One possibility is to consider string and quantum
corrections to the five-dimensional action. One would expect string
effects to play a role as the branes come near each other and quantum
effects to become important near a curvature singularity. The problem
of curing an ill-behaved collapse with string and quantum corrections
(including the question of matching conditions) in the ekpyrotic scenario is
in some ways reminiscent of the graceful exit problem in the pre-big bang model
\cite{Gasperini:1996, Brustein:1997, Foffa:1999, Cartier:1999}. In the
ekpyrotic senario the problem may seem more tractable, since one is approaching
the weak string coupling regime rather than the strong coupling regime as in
the pre-big bang model, as emphasised in \cite{Khoury:2001c}.

However, the problems of the ekpyrotic scenario and the pre-big bang model are
of different nature. In the pre-big bang case, the curvature singularity
appears from the equations of motion, so that it can in principle be avoided by
changing the action. In the ekpyrotic case, the conclusions on negative energy
density were made directly from the requirement of non-singularity without
recourse to the equations of motion apart from the junction conditions. So,
higher order curvature terms or string corrections to the
five-dimensional action can only help by changing the junction conditions.
Since the conclusions on negative energy density and pressure have
been drawn from singular contributions and the string coupling is posited to
vanish at the collision, string effects seem an unlikely remedy. 
Higher order curvature terms do in general change the junction conditions,
but the survival of some constraints on brane matter seems likely. This is
simply because though the near-collision metrics \re{asymptotick=1},
\re{asymptotickge2} include enough free parameters to account for the four
physical quantities (the energy densities and pressures on the two branes),
the parameters enter in a quite restricted manner.

There is always the possibility of going further with the dimensional lifting,
straight to the full eleven-dimensional string theory instead of the
effective five-dimensional field theory. However, five-dimensional brane
cosmologies have the merit of being relatively tractable and well-studied.
In particular, the treatment of perturbations has been under study \cite{Perturbations, Brax:2001b}, and may be applied to the five-dimensional picture of
the ekpyrotic scenario.

Before adding ingredients, one should be confident that the
extra complexity is really needed. The boundary brane collision was originally
introduced to solve the problem of collapse of the transverse direction,
something that was considered impossible for a bulk brane-boundary brane
collision \cite{Khoury:2001c, Khoury:2001d}. However, the conclusion that the
transverse direction contracts was based on a four-dimensional analysis
with the moduli space approximation. Note that the near-collision metrics
\re{asymptotick=1}, \re{asymptotickge2} are not of the moduli space form, see
\re{modulimetric}. It was already known that the moduli space approximation
cannot describe the post-ekpyrosis universe with brane matter
\cite{Kallosh:2001b, Enqvist:2001} (a similar result is well-known in
the Randall-Sundrum context \cite{Mohapatra:2000, Lesgourges:2000}), and now
we see that it cannot describe the pre-ekpyrosis universe either, at least in
the vicinity of the collision. It is then important to check whether the
conclusions in the bulk brane case regarding the collapse are borne out by the
full five-dimensional analysis. If the collapse problem turns out to be an
artifact of the moduli space approximation, one can then return to the
original proposal with the bulk brane and avoid the problems of boundary brane
collisions.

\section{Bulk brane collision}

We will now consider the original realisation of the ekpyrotic scenario with
a third brane in the bulk with the aim of checking the validity of the moduli
space approximation. We will assume that the moduli space approximation is
valid and see whether the results of the five-dimensional analysis of this
ansatz agree with those given by the four-dimensional effective theory.
Observations on the moduli space approximation in the context of the
five-dimensional theory have been previously made in \cite{Kallosh:2001b}.

\subsection{The moduli space approximation}

The metric of the BPS solution of five-dimensional heterotic M-theory with
minimal field content and three branes is \cite{Lukas, Khoury:2001a, Kallosh:2001b}
\bea \label{BPSmetric}
  ds^2 = -N^2 D(y) dt^2 + A^2 D(y)\sum^3_{j=1}(dx^j)^2 + B^2 D(y)^4 dy^2 \ ,
\eea

\noindent where $D(y)=\alpha y+C$ for $y<Y$ and $D(y)=(\alpha-\beta) y+C+\beta Y$ for $y>Y$ and $N, A, B, C$ and $Y$ are constants, with $Y$ being the
position of the bulk brane. The size of the Calabi-Yau threefold 
and the four-form field strength are given by
\bea
  e^{\phi(y)} &=& B D(y)^3 \el 
  {\cal F}_{0123y}(y) &=& - ( \alpha-\beta\theta(y-Y) ) A^3 N B^{-1} D(y)^{-2} \ .
\eea

The moduli space approximation consists of promoting the constants
$N, A, B, C$ and $Y$ to functions which depend on coordinates parallel to the
branes, which in the homogeneous and isotropic approximation means that they
depend only on time. Also, a potential for the modulus $Y$ is added to the
theory to support the movement in the space spanned by the moduli.

The metric of the moduli space approximation used as the basis of the
original realisation of the ekpyrotic scenario is thus
\bea \label{modulimetric}
  ds^2 = -N(t)^2 D(t,y) dt^2 + A(t)^2 D(t,y)\sum^3_{j=1}(dx^j)^2 + B(t)^2 D(t,y)^4 dy^2 \ ,
\eea

\noindent with
\bea \label{D}
  D(t,y) &=& \left\{ \begin{array}{cc}
  \alpha y + C(t) \ \qquad\qquad\qquad\qquad y\le Y(t) \el
  (\alpha-\beta) y + C(t) + \beta Y(t) \quad\quad y\ge Y(t)
\end{array} \right. \el
  &=& \alpha y + C(t) - \beta (y-Y(t)) \theta(y-Y(t)) \ ,
\eea

\noindent and the size of the Calabi-Yau threefold and the four-form field
strength are given by
\bea \label{moduliphiandF}
  e^{\phi(t,y)} &=& B(t) D(t,y)^3 \el
  {\cal F}_{0123y}(t,y) &=& - (\alpha-\beta\theta(y-Y(t)) A(t)^3 N(t) B(t)^{-1} D(t,y)^{-2} \ .
\eea

The bulk brane starts at the hidden brane, $Y=R$ and ends up at the visible
brane, $Y=0$, so that $\Ydot<0$.

In \cite{Khoury:2001a}, the moduli space approximation was substituted
into the action, which was then integrated over $y$ to obtain a
four-dimensional theory. The analysis was performed in the context of this
four-dimensional effective theory. This approximation was proposed to be valid
for slow evolution of the system. We will work directly with the
five-dimensional equations \re{eom} and \re{einstein}.

\subsection{Bulk brane movement and contraction}

The reason for replacing a bulk brane-boundary brane collision with a boundary
brane-boundary brane collision was that during the movement of the bulk brane
the transverse direction seemed to be collapsing
\cite{Khoury:2001a, Khoury:2001c}. Let us now check whether this result of the
four-dimensional effective theory is in agreement with the five-dimensional
equations.

We again have the requirement that energy should not flow away from spacetime:
\bea \label{flow}
  T_{\hat t\hat y}\bigg|_{y=y_i} &=& M_5^3 G_{\hat t\hat y}\bigg|_{y=y_i} \el
  &=& 3 M_5^3 \frac{1}{n b}\left( \frac{n'}{n}\frac{\adot}{a}+\frac{a'}{a}\frac{\bdot}{b}-\frac{\adot'}{a} \right)\bigg|_{y=y_i} \el
  &=& 0 \ .
\eea

Inserting the moduli space metric \re{modulimetric} into \re{flow}, we have
\bea
  T_{\hat t\hat y}\bigg|_{y=y_i} &\propto& \left(\frac{\Bdot}{B} + 3 \frac{\Ddot}{D} \right)\bigg|_{y=y_i} \el
  &=& 0 \ .
\eea

Inserting $D$ from \re{D} at $y=0$ and $y=R$, we have
\bea
  \frac{\Bdot}{B} &=& -3\frac{\beta\Ydot}{\beta Y + (\alpha-\beta)R} \el
  \frac{\Cdot}{C} &=& \frac{\beta\Ydot}{\beta Y + (\alpha-\beta)R} \ .
\eea

Integrating, we obtain
\bea \label{BandC}
  B(t) &=& B_0 (\beta Y(t) + (\alpha-\beta) R)^{-3} \el
  C(t) &=& C_0 (\beta Y(t) + (\alpha-\beta) R) \ ,
\eea 

\noindent where $B_0$ and $C_0$ are constants. As an aside, let us note that
in the approximation \mbox{$B=$ constant,} $C=$ constant used in
\cite{Khoury:2001a} the bulk brane cannot move at all. Also, in the boundary
brane case there is no bulk brane and hence no $Y$, so that the boundary
branes cannot move at all.

With \re{BandC} we can calculate the change in the size of the transverse
direction.
\bea \label{Ldot}
  \Ldot(t) &\equiv& \int_0^R dy\, \bdot(t,y) \el
  &=& \int_0^R dy ( \Bdot(t) D(t,y)^2 + 2 B(t)D(t,y)\Ddot(t,y) ) \el
  &=& -\frac{\beta \Ydot}{\beta Y + (\alpha-\beta) R} B R (\alpha Y + C)^2 \ .
\eea

Since $\Ydot$ is negative and $\beta$ is positive (and smaller than
$\vert\alpha\vert$), $\Ldot$ has the same sign as $\alpha$. In
\cite{Khoury:2001a} $\alpha$ was positive (and $-\alpha$, the tension of the
visible brane, negative), and the transverse direction was collapsing as the
bulk brane moved towards the visible brane, according to the four-dimensional
effective theory. We see that the five-dimensional equations lead to the
opposite conclusion: the transverse direction expands for $\alpha>0$.

Since the Newton's constant measured on a brane has the same sign
as the brane tension \cite{Mennim:2000, Enqvist:2001}, the tension of
the visible brane should be positive, $\alpha<0$. Then the transverse
direction is indicated to contract during bulk brane movement, so that
stabilisation would in general seem to be a problem. However, there is a
problem only if the moduli space approximation is valid, even in the
five-dimensional picture. At any rate, since the moduli space metric
\re{modulimetric} does not support brane matter and thus cannot describe the
post-ekpyrosis era, it is clearly not an adequate framework for
considering the issue.

\subsection{Bulk brane movement and the equations of motion}

After reviewing the collapse of the transverse direction, let us
consider bulk brane movement more generally. Since the brane interaction
has only been presented in the context of the four-dimensional effective field
theory, and it is non-trivial to see what it would
look like in the five-dimensional setting, we cannot solve the Einstein
equation directly. However, the moduli approximation is so constraining that
it is possible to obtain some results in spite of our ignorance.

\paragraph{The field equation of $\mathbf{\phi}$.}

Let us first assume that the brane interaction does not couple to the
modulus $\phi$, so that its equation of motion \re{eom} remains unaffected:
\bea
  \label{eomd} - n^{-2} \left[\phiddot + \left(-\frac{\ndot}{n}+3\frac{\adot}{a}+\frac{\bdot}{b}\right)\phidot\right] \el
  + b^{-2} \left[\phi'' + \left(\frac{n'}{n}+3\frac{a'}{a}-\frac{b'}{b}\right)\phi'\right] - \frac{3}{5!}e^{2\phi}{\cal F}_{ABCDE}{\cal F}^{ABCDE} &=& 0 \\
  \label{eomdeltad} \delta(y-y_i) ((-1)^{i+1}b^{-1}\phi'+3\, \alpha_i\, e^{-\phi}) &=& 0 \ .
\eea

The delta function part of the equation of motion, \re{eomdeltad}, is
satisfied automatically for the ansatz \re{modulimetric}, \re{moduliphiandF}.
The bulk part is not trivially satisfied and reads
\bea \label{modulieom}
  D^{-1} \left[ \frac{\Bddot}{B}+\frac{\Bdot}{B}\left( -\frac{\Ndot}{N}+3\frac{\Adot}{A} \right)\right] + 3 D^{-2}\left[ \Dddot + \Ddot\left( -\frac{\Ndot}{N}+3\frac{\Adot}{A}+2\frac{\Bdot}{B} \right)\right] + 6 D^{-3} \Ddot^2 = 0 \ .
\eea

The coefficient of each inverse power of $D$ in the above equation
has to vanish separately, for both $y<Y$ and $y>Y$. The $D^{-3}$ term 
then yields the result that $\Cdot=0$ and $\Ydot=0$. We see that unless the
brane interaction is coupled to $\phi$, the bulk brane
cannot move within the confines of the moduli space approximation. Even if the
brane interaction did contribute to the equation of motion, the results
$\Cdot=0$ and $\Ydot=0$ (and $\Cddot=0$, $\Yddot=0$) would still hold at the
collision, since the brane interaction vanishes at the collision. Let us note
that the brane interaction in \cite{Khoury:2001a} included only a delta
function part, so that the bulk brane would not move at all.

\paragraph{The Einstein equation.}

The conclusion that $\Cdot=0$, $\Ydot=0$, $\Cddot=0$ and $\Yddot=0$ at the
collision also follows from the Einstein equation. Even though we do not
know what the brane interaction is like, it is possible to extract some
information from the Einstein tensor due to the highly restrictive form
of the moduli space metric. Namely, since
\bea
  G_{AB} &=& \frac{1}{M_5^3} T_{AB} \el 
  &=& \frac{1}{M_5^3} T_{AB}(\phi) + \frac{1}{M_5^3} T_{AB}(BI) \ ,
\eea

\noindent the energy-momentum tensor of the brane interaction is given by
\bea
  \frac{1}{M_5^3} T_{AB}(BI) &=& G_{AB} - \frac{1}{M_5^3} T_{AB}(\phi) \ . 
\eea

Inserting the metric and the fields $\phi$ and ${\cal F}_{ABCDE}$ in the
moduli space approximation \re{modulimetric}, \re{moduliphiandF} into the
Einstein equation \re{einstein}, we obtain
\bea
  \frac{1}{M_5^3} T^t_{\ t}(BI) &=& -3 D^{-1} N^{-2} \left( \frac{\Adot^2}{A^2} + \frac{\Adot}{A}\frac{\Bdot}{B} - \frac{1}{12}\frac{\Bdot^2}{B^2} \right) - 9 D^{-2} N^{-2}\Ddot\frac{\Adot}{A} - \frac{3}{2} D^{-3} N^{-2}\Ddot^2 \el
  \frac{1}{M_5^3} T^j_{\ j}(BI) &=& -D^{-1} N^{-2} \left( 2\frac{\Addot}{A} + \frac{\Bddot}{B} + \frac{\Adot^2}{A^2} - 2\frac{\Adot}{A}\frac{\Ndot}{N} + 2\frac{\Adot}{A}\frac{\Bdot}{B} - \frac{\Ndot}{N}\frac{\Bdot}{B} + \frac{1}{4}\frac{\Bdot^2}{B^2} \right) \el
  && - 3 D^{-2} N^{-2} \left( \Dddot - \frac{\Ndot}{N}\Ddot + 2\frac{\Adot}{A}\Ddot + 2\frac{\Bdot}{B}\Ddot \right) - \frac{9}{2} D^{-3} N^{-2}\Ddot^2 \el
  \frac{1}{M_5^3} T^y_{\ y}(BI) &=& -3 D^{-1} N^{-2} \left( \frac{\Addot}{A} + \frac{\Adot^2}{A^2} - \frac{\Adot}{A}\frac{\Ndot}{N} + \frac{1}{12}\frac{\Bdot^2}{B^2} \right) \el
  && -\frac{3}{2} D^{-2} N^{-2} \left( \Dddot - \frac{\Ndot}{N}\Ddot + 3\frac{\Adot}{A}\Ddot + \frac{\Bdot}{B}\Ddot \right) - \frac{3}{2} D^{-3} N^{-2}\Ddot^2 \el
  \frac{1}{M_5^3} T_{ty}(BI) &=& 0 \ .
\eea

Let us recall that the brane interaction is posited to vanish as the bulk
approaches the visible brane, $Y\rightarrow0$. Every coefficient of an inverse
power of $D$ has to vanish separately, so that near the collision
$\Cdot, \Ydot, \Yddot$ and $\Cddot$ all approach zero.

\subsection{Summary of bulk brane collisions}

We have shown that in the moduli space approximation the bulk brane
cannot move unless the bulk part of the brane interaction is coupled to the
size of the Calabi-Yau threefold. Even if the bulk brane moved, its
velocity (and thus kinetic energy) would vanish at the collision, leading to
zero ekpyrotic temperature, according to the formulae of \cite{Khoury:2001a}.
Furthermore, should the bulk brane move towards the visible brane,
the transverse direction would (for a negative tension visible brane)
expand as opposed to contracting. As an aside, we have noted that in the
moduli space approximation, the boundary branes cannot move at all without
the presence of a bulk brane. Further, it is well-known that metrics of the
moduli space form do not support brane matter and thus cannot describe the
post-ekpyrosis era.

The above points and especially their frequent contradiction with the results
of the four-dimensional analysis raises serious doubts about the validity of
the moduli space approximation and the four-dimensional effective theory based
on this approximation.

\section{Conclusion}

\paragraph{Implications for the ekpyrotic scenario.}

We have derived the five-dimensional metrics that describe non-singular
boundary brane collisions in general relativity under the assumptions of
BPS embedding of the branes, and homogeneity, isotropy and flatness in
the spatial directions parallel to the branes. These metrics imply that
branes contain exotic matter after the collision. Negative energy
density can possibly be avoided by turning on additional fields, and negative
pressure possibly by adding quantum corrections to the five-dimensional action
or by considering the actual string theory instead of the effective
five-dimensional field theory. However, since the moduli
space result that the fifth dimension collapses when a bulk brane travels
across it does not seem sound, the simplest way to bypass the problems might
be to go back to the original scenario with the bulk brane. The outlook would
then be to find what the brane interaction looks like in five dimensions,
solve the Einstein equation and the field equations for the background, and
then do the perturbation analysis, building on existing methods for brane
cosmologies in five dimensions \cite{Perturbations, Brax:2001b}.

\paragraph{Comments on the ``cyclic model of the universe''.}

Our results have some bearing on the recently proposed ``cyclic model of
the universe'' \cite{Steinhardt:2001a, Steinhardt:2001b}. This scenario,
also based on heterotic M-theory, proposes that ekpyrosis occurs at regular
intervals, with inflation serving to empty the branes between collisions.
The idea is quite interesting, with the unification of late-time acceleration
with primordial inflation being especially appealing. However, the treatment
of the hidden dimensions seems to have some flaws.

First, the cyclic model has been presented within the framework of a
four-dimensional effective theory, with the interbrane distance appearing as
a scalar field in the Hubble law on the brane. Though the theory in five
dimensions is evidently not the same as the ekpyrotic moduli space
approximation, any theory based on a factorisable metric to be
integrated over the fifth dimension is likely to suffer from similar problems.
In particular, the exact five-dimensional equations show that the while the
volume of the Calabi-Yau space can have a significant effect on the Hubble law
on the brane, the interbrane distance makes no direct contribution
\cite{Enqvist:2001}. In particular, it does not appear as a scalar field. This
is a general feature of brane cosmologies, where matter is localised on a
slice of spacetime as opposed to being spread across a hidden dimension
\cite{Shiromizu:1999}.

Second, the Calabi-Yau space is kept fixed and ignored. However, this
violates the equation of motion of $\phi$ near the collision, \re{eomb},
\re{eomdeltab}. Even if this were not the case, the energy (and the pressures)
associated with a constant breathing modulus $\phi$ would grow without bound at
the approach to the collision, as we see from \re{emtensortt}--\re{emtensoryy}.
If the Calabi-Yau volume is kept fixed only at the position of the visible
brane, there is no obvious divergence or contradiction with the equation of
motion. However, the brane collision will then either produce negative
energy density or be singular, as we have seen in section 3.

Third, it has been proposed that near the collision, spacetime can be treated
as flat, as argued in \cite{Khoury:2001c}. However, brane tension necessarily
implies that curvature cannot be neglected, as we have seen in section 3.
Further, the near-collision metric is not even factorisable as in the
Kaluza-Klein approach used in \cite{Steinhardt:2001a, Steinhardt:2001b}.

While these problems seem integral to the proposal presented in
\cite{Steinhardt:2001a, Steinhardt:2001b}, the interesting ideas of the
``cyclic model of the universe'' will hopefully be realised in more thorough
explorations of brane cosmology.

\section*{Acknowledgements}

I thank Kari Enqvist and Esko Keski-Vakkuri for guidance, Martin Sloth for
comments on the manuscript and Fawad Hassan, Andrei Linde, Burt Ovrut,
Paul Steinhardt and Riccardo Sturani for discussions. Thanks are also due to
Andrew Liddle for encouragement. The research has been partially supported by
grants from the Jenny and Antti Wihuri Foundation and the Magnus Ehrnrooth
Foundation.

\end{document}